\documentclass[11pt]{article}
\usepackage{epsfig}

\newcommand{\be}{\begin{eqnarray}}
\newcommand{\ee}{\end{eqnarray}}

\def\ll#1{\left#1}
\def\r#1{\right#1}
\def\fr{\frac{1}{2}}

\def\mref#1{(\ref{#1})}

\def\p{\partial}

\def\bd{\begin{displaymath}}
\def\ed{\end{displaymath}}

\def\ba#1{\begin{array}{#1}}
\def\ea{\end{array}}
\def\nn{\nonumber}
\newfont{\Bbb}{msbm10 scaled 1200}

\begin{document}
\pagestyle{empty}

\begin{center}

{\LARGE\bf On the Consistency of Twisted Gauge Theory
\footnote[1]{Supported by the grant no 1P03B02128 of the Polish Ministry of Science}\\[0.5cm]}

\vskip 12pt

{\large {\bf Stefan Giller${\dag}$\\Cezary Gonera$\ddag$\\Piotr Kosi\'nski$\ddag$\\
Pawe{\l} Ma\'slanka$\ddag$}}

\vskip 3pt

$^\dag$Jan Dlugosz Academy in Czestochowa\\
Institute of Physics\\ul. Armii Krajowej 13/15, 42-200 Czestochowa, Poland\\
e-mail: sgiller@uni.lodz.pl\\
 $\ddag$Theoretical Physics Department II, University of {\L}\'od\'z,\\
Pomorska 149/153, 90-236 {\L}\'od\'z, Poland\\ e-mail:
cgonera@uni.lodz.pl \\
pkosinsk@uni.lodz.pl\\
pmaslan@uni.lodz.pl
\end{center}
\vspace{10pt}
\begin{abstract}
It is argued that the twisted gauge theory is consistent provided it exhibits also the standard noncommutative gauge
symmetry.
\end{abstract}

\newpage

\pagestyle{plain}

\setcounter{page}{1}

In the recent years the noncommutative field theories \cite{1,2,3} have gained much attention. This is in part due to
the fact that they emerge as a specific limit of string theory and, being much simpler than the original string theory,
yet keep some of its properties \cite{4,5}. Moreover, they can be viewed as an attempt to describe the space-time at
very short distances where the very idea of space-time as a smooth manifold may lack any meaning.

Due to the crucial role played by gauge theories it is natural to ask whether the idea of gauge symmetry can be
generalized to the case of the field theory on noncommutative space-time. When trying to extend in this direction the
notion of gauge transformations one is faced with some serious problems because for many classical groups the
modification consisting in replacing the product with star product leads to the transformations which do not close to
form a group. These problems lead to some no-go theorems restricting the group and representing content of
noncommutative gauge theories \cite{6}. A number of more or less sophisticated proposals has been made to avoid the
troubles with the idea of the localization of symmetry on noncommutative space-time \cite{7,8,9}.

 The most flexible
one seems to be the approach based on the idea of twisted gauge symmetry \cite{10,11,12,13}. Putting aside all formal
details one can summarize the notion of twisted gauge transformations as follows: in order to compute the gauge variation
of star-product monomial in fields and their derivatives one applies the Leibnitz rule but always keeping the gauge
parameter function leftmost. It is easily seen that in this case any gauge group can be closed and the twisted gauge
invariance imposes rather mild restrictions on the form of the action.
The idea of twisted gauge transformations seems to provide a general method of introducing the gauge symmetry for
field theories on noncommutative space-time. However, according to Milton Friedman, there are no lunches for free. This
observation applies here as well. First it has been argued the the approach based on twisted gauge transformations is
in contradiction with the very concept of gauge fields \cite{14}.

Another serious arguments are given in the paper of Alvarez-Gaume {\it et al}. \cite{15}. They noticed that the modified
Leibnitz rule (due to the fact that the gauge parameter function is kept leftmost) means that the twisted symmetry is
not a physical symmetry in the usual sense and it is not clear whether the Noether theorem can be derived. For the case
of twisted $U(N)$ gauge theory the action is also invariant under the standard noncommutative $U(N)$ symmetry and this
could be the reason for the existence of conserved current. This observation may be linked with the one made in Refs.
\cite{11,12}. It has been observed \cite{11,12} that, in spite of the fact that one can gauge any group, when writting the invariant
action one must admit that the gauge potential takes values in larger algebra than the Lie algebra of the gauge group.
This is not only because the field strength takes necessarily values in enveloping algebra but is also somehow necessary for
the consistency of field equations \cite{12}. In fact, it appears that in all examples considered so far the range of
values of the gauge potential is such that the action invariant under twisted gauge transformations is also
invariant under standard noncommutative gauge symmetry; the latter enforces the consistency of current conservation and
field equations.

In the present note we argue that this is generally the case. Namely, if one considers the theory invariant under twisted
gauge symmetry then, in general, the current conservation implies some constraints which are not automatically fulfiled
by virtue of field equations; more specifically, these constraints follow by taking the divergence of field equtaions.
If one demands no such constraints exist, the range of gauge potentials must be enlarged so that the action exhibits
also standard noncommutative gauge symmetry. Therefore, the twisted gauge theory, if consistent, posses a custodial
standard symmetry enforcing current conservation.

For gauge theories on commutative space-time the local gauge invariance allows to derive the set of identities \cite{16}
implying that the divergencies of Euler-Lagrange expressions vanish by virtue of field equations. We explain why this is
also the case for noncommutative gauge symmetry but not for the twisted one. Roughly speaking, this is because the variation
of action under standard noncommutative gauge symmetry is computed in the same way as the variation leading to the field equtions.
More detailed discussion is given in the final part of the paper.

We start with the following general example.

Let $G$ be some compact Lie group with the Lie algebra $g$. We want to construct the gauge theory on noncommutative
space-time based on twisted gauge symmetry with $G$ playing the role of gauge group. To this end we select some $N$-dimensional
unitary irreducible representation of $G$ (and the corresponding irreducible representation of $g$, denoted also by $g$)
and assume that the gauge potential takes its values in $g$. The Lie algebra $g$ is a subalgebra of $u(N)$. The latter
viewed as linear space is equipped with the scalar product $(A,B)\equiv Tr(AB)$ invariant under the adjoint action of $U(N)$.
Let $g_{\perp}$ be the orthogonal complement of $g$ in $u(N)$ and let $B_i,\;i=1,...,a$, be an orthonormal bases in
$g_{\perp}$. The subspace $g$ can be now characterized by the condition: $A\in g\Longleftrightarrow
Tr(AB_i)=0,\;i=1,...,a$.

The gauge theory with twisted $G$-symmetry is now formulated with help of Lagrange multiplier method. To this end we assume
that the gauge potential $A_\mu$ {\it a priori} takes its values in $U(N)$ and consider the action:
\be
S=-\frac{1}{4}\int d^4xTr(F_{\mu\nu}*F_{\mu\nu})+\int d^4x\rho_i^\mu Tr(A_\mu B_i)
\label{1}
\ee
where
\be
F_{\mu\nu}\equiv \p_\mu A_\nu-\p_\nu A_\mu -i[A_\mu,A_\nu]_*
\label{2}
\ee
while $\rho_i^\mu(x)$ are the Lagrange multipliers.

Eq. \mref{1} implies the following equations of motion
\be
Tr(A_\mu B_i)=0\nn\\
\p_\mu F^{\mu\nu}-i[A_\mu,F^{\mu\nu}]_*+\rho_i^\nu B_i=0
\label{3}
\ee

Eliminating the Lagrange multipliers one obtains
\be
\p_\mu F^{\mu\nu}-i[A_\mu,F^{\mu\nu}]_*-Tr(B_i(\p_\mu F^{\mu\nu}-i[A_\mu,F^{\mu\nu}]_*))B_i=0
\label{4}
\ee

When projected on $g$ the above eqution yields standard Yang-Mills equation
\be
\ll.\p_\mu F^{\mu\nu}-i[A_\mu,F^{\mu\nu}]_*\r|_g=0
\label{5}
\ee

However, one must keep in mind that even if the gauge potential takes its values in $g$ this is no longer the case for
the field strength tensor $F_{\mu\nu}$. Therefore, $\p_\mu F^{\mu\nu}-i[A_\mu,F^{\mu\nu}]_*$ can have a nonvanishing
component in $g_\perp$. In fact, eq.\mref{4} is just the statement that $\p_\mu F^{\mu\nu}-i[A_\mu,F^{\mu\nu}]_*\in g_\perp$.

In the case of commutative space-time $\p_\mu F^{\mu\nu}-i[A_\mu,F^{\mu\nu}]_*\in g$ by construction so the field equations
imply $\p_\mu F^{\mu\nu}-i[A_\mu,F^{\mu\nu}]_*=0$. In the standard (i.e. not twisted) approach to gauge theory on
noncommutative space-time $g$ must basically be $u(N)$ and $g_\perp=0$; again one arrives at standard form of field
equations.

Having obtained the field equations we can ask whether they are consistent with the requirement of current
conservation. This seems to be by far not obvious. In fact, taking the divergence of eq.\mref{4} results in the
following additional condition:
\be
Tr\ll(\Gamma\ll[A_\nu,Tr(B_i(\p_\mu F^{\mu\nu}-i[A_\mu,F^{\mu\nu}]_*))B_i\r]_*\r)=0
\label{6}
\ee
for any $\Gamma\in g$. We would like to know under which circumstances eq.\mref{6} does not provide further constraints on
gauge fields. Let us note that the star commutator under the outer trace consists of two pieces, one proportional to the
matrix commutator and the second one involving matrix anticommutator. Now by virtue of field equations \mref{2} $A_\mu$
belongs to $g$. It is straightfarward to check that $[\Gamma',B_i]\in g$ for any $\Gamma'\in g$. Indeed, if $U\in G$ then
$U^+\Gamma'U\in g$ and $Tr(\Gamma'UB_iU^+)=Tr(U^+\Gamma'UB_i)=0$ so that $UB_iU^+\in g_\perp$. Therefore, the part
involving commutator gives no constraint. The piece containing the anticommutator does not product any constraint only
provided $\{\Gamma',B_i\}\in g_\perp$ for $\Gamma'\in g$. This together with $[\Gamma',B_i]\in g_\perp$ implies that both
$\Gamma'B_i\in g_\perp$ and $B_i\Gamma'\in g_\perp$ so that $\Sigma B_i\in g_\perp$ and $B_i\Sigma\in g_\perp$ for each
$\Sigma$ belonging to the enveloping algebra of $g$. By assumption $g$ is irreducible; the Burnside theorem implies then
that $B_i$ span two-sided ideal in the algebra of all $N\times N$ matrices. However, by virtue of Wedderburn theorem
such an ideal, if proper (only then there is some gauge symmetry at all) must be zero. Then $g=u(N)$ and we are dealing
with the theory exhibiting standard noncommutative gauge symmetry.

Consider some more specific examples of the situation described above. In Ref\cite{11} the $SU(2)$ twisted gauge theory
is considered. However, it is admitted from the very beginning, that the gauge field takes its value in $U(2)$ algebra
and that the resulting theory is invariant under the standard $U(2)$ noncommutative gauge symmetry which guarantees the
existence of the conserved current. On the other hand, if we assume that $A_\mu$ belongs to the $SU(2)$ algebra only the
second equation (4.29) of Ref.\cite{11} survives and, additionally, $B_\mu=0$. One easily checks that by taking the
divergence of this equation one obtains a new constraint of the form \mref{6}.

As a second example take $G=SO(N)$ in the defining representation. Then $g$ is the set of imaginary antisymmetric
$N\times N$ matrices while $g_\perp$ - the set of the real symmetric ones. Again one can write the Lagrangian containing
gauge fields taking their values in $SO(N)$. Takin the divegence of the resulting field equations and separating
symmetric and antisymmetric parts one arrives at the constraint of the general form \mref{6}.

Finally, let us consider the simple example of gauge theory with matter fields \cite{12}. The gauge group is $U(1)$ and
the matter fields form the multiplet tranforming as follows
\be
\delta_\alpha\Psi=i\alpha {\bf Q}\Psi,\;\;\;\;\;\;{\bf Q}=\ll[\begin{array}{cc}
                                              1&0\\
                                              0&-1
                                              \ea\r]
\label{7}
\ee

The gauge field takes its values in the algebra spanned by  $\bf I$ and $\bf Q$
\be
{\bf A}_\mu=A_\mu{\bf I}+B_\mu{\bf Q}
\label{8}
\ee
and the Lagrangian reads
\be
L=-\frac{1}{4}Tr({\bf F}^{\mu\nu}*{\bf F}_{\mu\nu})+{\bar\Psi}*\gamma^\mu(i\p_\mu+A_\mu*)\Psi-m{\bar\Psi}*\Psi
\label{9}
\ee
where
\be
{\bf F}_{\mu\nu}\equiv \p_\mu{\bf A}_\nu-\p_\nu{\bf A}_\mu -i[{\bf A}_\mu,{\bf A}_\nu]_*
\label{10}
\ee

In order to reveal the content of the theory we define two projectors
\be
{\bf P}_\pm\equiv\fr({\bf I}\pm{\bf Q})
\label{11}
\ee
and the corresponding decomposition of the gauge field
\be
{\bf A}_\mu=A_\mu^+{\bf P}_++A_\mu^-{\bf P}_-,\;\;\;\;\;\;A_\mu^\pm\equiv A_\mu\pm B_\mu
\label{12}
\ee
and the matter field
\be
\Psi_\pm\equiv{\bf P}_\pm\Psi,\;\;\;\;\;\Psi=\Psi_++\Psi_-
\label{13}
\ee

Then
\be
{\bf F}_{\mu\nu}= F_{\mu\nu}^+{\bf P}_+ + F_{\mu\nu}^-{\bf P}_-
\label{14}
\ee
with
\be
F_{\mu\nu}^\pm\equiv \p_\mu A_\nu^\pm-\p_\nu A_\mu^\pm -i[A_\mu^\pm,A_\nu^\pm]_*\nn\\
\nn\\
(\p_\mu-iA_\mu*)\Psi=(\p_\mu-iA_\mu^+*)\Psi_++(\p_\mu-iA_\mu^-*)\Psi_-
\label{15}
\ee

Therefore, the Lagrangian splits into two independent parts
\be
L=-\frac{1}{4}Tr(F^{+\mu\nu}*F_{\mu\nu}^+)+{\bar\Psi}_+*\gamma^\mu(i\p_\mu+A_\mu^+*)\Psi_+-m{\bar\Psi}_+*\Psi_+-\nn\\
\frac{1}{4}Tr(F^{-\mu\nu}*F_{\mu\nu}^-)+{\bar\Psi}_-*\gamma^\mu(i\p_\mu+A_\mu^-*)\Psi_--m{\bar\Psi}_-*\Psi_-\nn\\
\label{16}
\ee

The "$\pm$" degrees of freedom decouple and both pieces exhibit standard (noncommutative) $U(1)$ symmetry.

The important point here is that the gauge potential is a matrix acting in representation space of matter fields.
However, consistency (i.e. the absence of further constraints) demands that $A_\mu$ takes its values in the enveloping
algebra of the Lie algebra under consideration. Therefore, the gauge potential is the set of {\underline two}
independent fields and the theory describes two idependent standard noncommutative $U(1)$-symmetric pieces.

On the other hand, if we insist on using the gauge potential belonging to the Lie algebra $U(1)$ (i.e. consisting of
one field) then either all fields have the same charge (i.e. the enveloping algebra coincides with the Lie algebra of
$U(1)$) or there is additional constraint\cite{12}.

In all the above examples in order to avoid additional constraints following from the field equations one has to extend
the definition of gauge potential in such a way that the resulting theory exhibits the standard noncommutative gauge
symmetry. It seems that just the very existence of the latter enforces consistency.

In order to put more light on the problem let us remind the second Noether theorem in th commutative case\cite{16}.
Consider the field theory described by the Lagrangian $L(\Phi,\p_mu\Phi)$ carrying the following local symmetry
\be
\delta\Phi^i(x)=a_k^i(\Phi;x)\epsilon^k(x)+b_k^{i\mu}(\phi;x)\p_\mu\epsilon^k(x)
\label{17}
\ee

Then the invariance of $L$ implies the following identities
\be
a_k^i\frac{\delta L}{\delta\Phi^i}+\p_\mu\ll(a_k^i\frac{\p L}{\p\ll(\p_\mu\Phi^i\r)}\r)\equiv 0\nn\\
b_k^{i\mu}\frac{\delta L}{\delta\Phi^i}+a_k^i\frac{\p L}{\p\ll(\p_\mu\Phi^i\r)}+
\p_\nu\ll(b_k^{i\mu}\frac{\p L}{\p\ll(\p_\mu\Phi^i\r)}\r)\equiv 0\nn\\
b_k^{i\nu}\frac{\p L}{\p\ll(\p_\mu\Phi^i\r)}+b_k^{i\mu}\frac{\p L}{\p\ll(\p_\nu\Phi^i\r)}\equiv 0
\label{18}
\ee
where
\[\frac{\delta L}{\delta\Phi^i}\equiv \frac{\p L}{\p\Phi^i}-\p_\mu\ll(\frac{\p L}{\p\ll(\p_\mu\Phi^i\r)}\r)\]
is the Euler -- Lagrange expression.

The first identity expresses the conservation of the current
\be
j_k^\mu \equiv a_k^i\frac{\p L}{\p\ll(\p_\mu\Phi^i\r)}
\label{19}
\ee
and is valid even if $L$ exhibits merely global symmetry. The remaining identities hold only for local symmetries. They
imply that the field equations are in fact dependent. Indeed, one easily derives the following Bianchi identities
\be
a_k^i\frac{\delta L}{\delta\Phi^i}-\p_\mu\ll(b_k^{i\mu}\frac{\delta L}{\delta\Phi^i}\r)=0
\label{20}
\ee

In the particular case of pure gauge theory eqs.\mref{14},\mref{18} take the following form (with $c_{bc}^a$ being the
structure constants)
\be
c_{bc}^a\ll(A_\mu^c\ll(\frac{\p L}{\p A_\mu^a}-\p_\nu\ll(\frac{\p L}{\p\ll(\p_\nu A_\mu^a\r)}\r)\r)+
\p_\nu\ll(A_\mu^c\frac{\p L}{\p\ll(\p_\nu A_\mu^a\r)}\r)\r)=0\nn\\
\frac{\p L}{\p A_\mu^a}-\p_\nu\ll(\frac{\p L}{\p\ll(\p_\nu A_\mu^a\r)}\r)-
c_{ac}^bA_\nu^c\frac{\p L}{\p\ll(\p_\mu A_\nu^b\r)} + \p_\nu\ll(\frac{\p L}{\p\ll(\p_\nu A_\mu^a\r)}\r)=0\nn\\
\frac{\p L}{\p\ll(\p_\nu A_\mu^a\r)}+\frac{\p L}{\p\ll(\p_\mu A_\nu^a\r)}=0
\label{21}
\ee
while \mref{20} corresponds to
\be
c_{bc}^aA_\mu^c\frac{\delta L}{\delta A_\mu^a}-\p_\mu\ll(\frac{\delta L}{\delta A_\mu^b}\r)=0
\label{22}
\ee

We see from eq.\mref{22} that no new relation (constraint) follows by taking the divergence of field equations.

In the case noncommutative field theories gauge symmetry also implies the set of identities. Let us consider first the
standard noncommutative gauge theory. In order to answer the question whether there exists some counterpart of Noether
identities \mref{21} let us recall the way the field equations are obtained from variational principle. One computes
the variation of the Langrangian by applying Leibnitz rule to each monomial entering the latter and making cyclic permutations
to bring the field variation say leftmost. Consequently, the field equations can be written in standard form provided the
derivatives $\frac{\p}{\p A_\mu^a}$, $\frac{\p}{\p\ll(\p_\nu A_\mu^a\r)}$ are defined as follows: in order to apply the
relevant derivatives to some monomial in gauge fields and their derivatives one uses the Leibnitz rule and then makes in
each term the cyclic permutation to bring the derivative leftmost. On the other hand, similar situation is encountered when
considering the standard noncommutative gauge symmetry. Again, one considers the variation of the Lagrangian implied, via
the Leibnitz rule, by the variation of gauge fields under the gauge transformation. The variation of action is obtained
by varying the Lagrangian under the integral so one can make cyclic transposition to bring the gauge parameter leftmost.
The result is that the coefficients behind gauge parameter and its first and second derivatives vanish. In this way we
obtain a set of identities in which derivatives $\frac{\p}{\p A_\mu^a}$, $\frac{\p}{\p\ll(\p_\nu A_\mu^a\r)}$ have the same
meaning as in the case of field equations. These identities have therefore very similar form to their commutative
counterparts. In particular, they also imply that no further constraint follows by taking the divergence of field
equations.

Let us now turn to twisted gauge symmetry. The formal definition of twisted gauge transformation can be summarized as
follows: the variation of arbitrary monomial in gauge fields and their derivatives is calculating according to the
Leibnitz rule but with the gauge parameter put leftmost (no cyclic permutation). The derivatives
$\frac{\p}{\p A_\mu^a}$, $\frac{\p}{\p\ll(\p_\nu A_\mu^a\r)}$ entering the identities expressing the invariance under
the twisted gauge symmetry are therefore defined by Leibnitz rule (without cyclic permutation). In spite of the fact that
they appear in the combination resembling Euler -- Lagrange expressions such combinations do not vanish by virtue of the
field equations due to the difference in the definition of derivatives. As a result, the modified Noether identities
do not guarantee the consistency of field equations. This seems to make the notion of twisted gauge symmetry less
interesting.

The structure of twisted gauge theories  has been also analysed in two recent papers \cite{17}, \cite{18}

\end{document}